\documentclass[twocolumn,trackchanges,lineno]{aastex6}
\usepackage{amsmath}
\usepackage{amssymb}
\usepackage{amsthm}
\usepackage{natbib,aasdefs,url,bm}
\usepackage{array}
\usepackage{float}
\usepackage{graphicx}
\usepackage{subfigure}
\usepackage{color}

\newcommand{\msun}{\rm M_{\odot}}
\newcommand{\mdotbh}{\dot M_{\rm B\mathchar`-H}}

\begin{document}

\title{Outflow-Driven Transients from the Birth of Binary Black Holes II:\\ Primary-Induced Accretion Transients}

\author{Shigeo S. Kimura\altaffilmark{1,2,3}, Kohta Murase\altaffilmark{1,2,3,4}, and Peter M\'{e}sz\'{a}ros\altaffilmark{1,2,3}}
\altaffiltext{1}{Department of Physics, Pennsylvania State University, University Park, Pennsylvania 16802, USA}
\altaffiltext{2}{Department of Astronomy \& Astrophysics, Pennsylvania State UNiversity, University Park, Pennsylvania 16802, USA}
\altaffiltext{3}{Center for Particle and Gravitational Astrophysics, Pennsylvania State University, University Park, Pennsylvania 16802, USA}
\altaffiltext{4}{Yukawa Institute for Theoretical Physics, Kyoto, Kyoto 606-8502, Japan}
\begin{abstract}
We discuss the electromagnetic radiation from newborn binary black holes (BBHs). 
As a consequence of the evolution of massive stellar binaries, a binary consisting of a primary black hole (BH) and a secondary Wolf-Rayet star is expected as a BBH progenitor system. We investigate optical transients from the birth of BBHs powered by the Bondi-Hoyle-Lyttleton accretion onto the primary BH, which occur $\sim1-10$~Gyr earlier than gravitational wave signals at the BH-BH merger.
When the secondary massive star collapses into a BH, it may eject a fraction of its outer material and may form a disk around the primary BH and induces a powerful disk wind. 
These primary-induced winds can lead to optical transients with a kinetic energy of $\sim10^{47}$ -- $3\times10^{48}$~erg, an ejecta velocity of $10^8$ -- $10^9\rm~cm~s^{-1}$, a duration of a few days, and an absolute magnitude ranging from about $-11$ to $-14$.
The light curves and late-time spectra of these transients are distinctive from those of ordinary supernovae, and detection of this type of transient is possible by future optical transient surveys if the event rate of this transient is comparable to the merger rate of BBHs.
This paper focuses on the emissions from disk-driven transients induced by the primary BH, different from the first paper that focuses on wind-driven transients from the tidally-locked secondary massive star.
\end{abstract}


\section{Introduction} \label{sec:intro}

The detections of gravitational waves (GWs) by the advanced Laser Interferometer Gravitational-wave Observatory \citep[LIGO;][]{LIGO16e,LIGO16a,LIGO17a} bring us a major new mystery of astrophysics: the formation process of binary black holes (BBHs) of $\sim30\rm\msun$. The evolution of isolated massive stellar binaries \citep[e.g.,][]{TY93a,KIH14a,MLP16a} is a straight-forward scenario, although other possibilities are also actively discussed, such as binaries of primordial black holes \citep[e.g.,][]{NST97a} and dynamical formation in dense stellar clusters \citep[e.g.,][]{SH93a,FTM17a}. The mass, spin, and redshift distributions of merging BBHs, which will be provided in future GW observations, are useful to distinguish the formation scenario of BBHs \citep[e.g.,][]{KZK16a}.

Another way to probe the environment of BBH formation is searching for electromagnetic radiations from newly born BBHs. 
These signals are not coincident with GW signals, since the merger events of BBHs typically take place $\sim0.1-10$ Gyr after their formation. 
According to typical binary evolution scenarios, BBHs are formed through the gravitational collapse of a secondary Wolf Rayet (WR) star in WR-BH binaries \citep[e.g.,][]{BHB16a}. There are two possibilities following the collapse, depending on the spin of the WR. 
The spin of the WR is determined by the tidal synchronization, whose time scale is estimated to be $t_{\rm TL}\sim 10^7 (t_{\rm mer}/1\rm~Gyr)^{17/8}~yr$, where $t_{\rm mer}=5c^5a^4/(512G^3M_*^3)$ is the GW inspiral time ($M_*$ is the primary mass and $a$ is the binary separation), and we assume $q=1$ for simplicity \citep{Zah77a,Tas87a,KZK16a}. 
This timescale is very sensitive to $M_*$ and $a$, so that both tidally locked and unlocked systems are possible.

\begin{figure*}[tbp]
\begin{center}
\includegraphics[width=\linewidth]{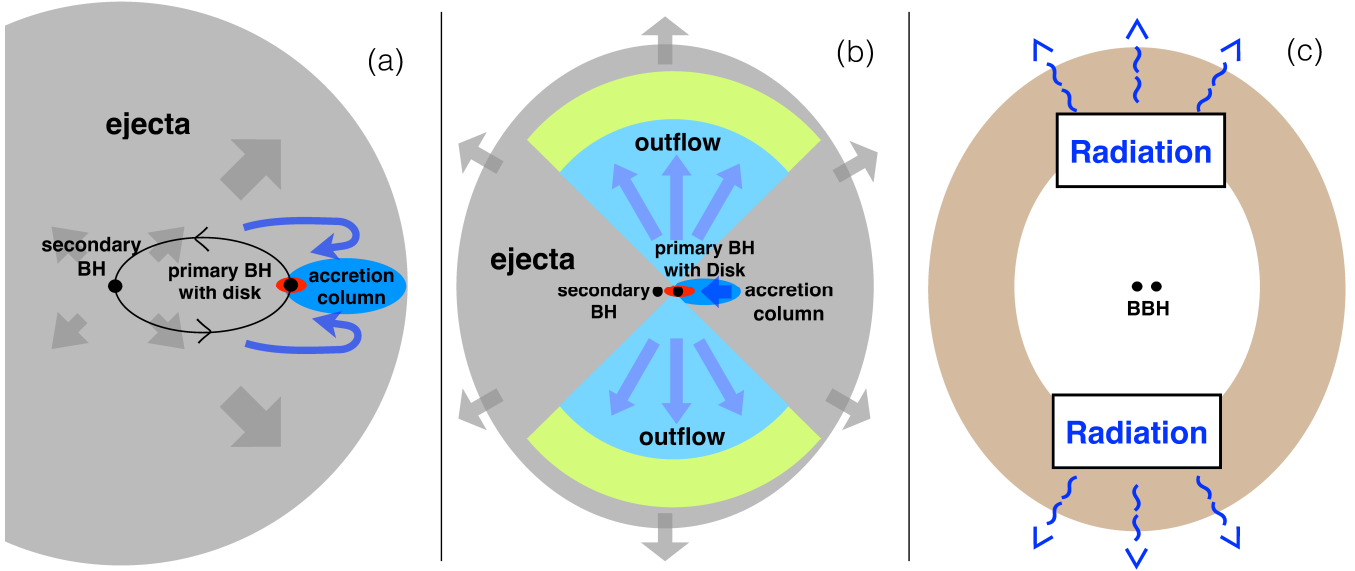}
\caption{Schematic picture of the primary-induced accretion transients considered in this work. 
 (a) At the birth of a BBH, a small amount of material is ejected by the secondary BH. 
It accretes onto the primary BH, forming a disk around the latter. 
(b) A disk-driven wind is launched. 
(c) Thermal radiation diffusively escape from the ejecta, 
while a fraction of the ejecta falls back onto the primary BH. }
    \label{fig:schematic}
   \end{center}
  \end{figure*}

For $M_*\sim10\rm~\msun$ and $a\sim10^{12}$ cm, the synchronization timescale is estimated to be $t_{\rm TL}\sim1.2\times10^8$~yr, which is much longer than the typical evolution time of massive stars ($\sim10^6$ yr).
In this case, the spin of the secondary is expected to be so slow that the secondary star can collapse directly to a BH
\footnote{Even if the binary separation is very small, both the centrifugal and Coriolis forces do not affect the disk formation process following the collapse of the WR star.}.
However, even in such cases, the sudden gravitational potential change due to neutrino loss may lead to a weak explosion~\citep{Nad80a}. 
Although the explosion itself is dim, brighter transients may be caused by the primary BH if the binary separation is sufficiently small. 
A fraction of the ejecta from the weak explosion should be gravitationally captured by the primary BH via Bondi-Hoyle-Littleton accretion \citep{HL39a,Bon52a}. 
This accreting material will form a disk around the primary, 
which results in a powerful wind\footnote{We use ``wind'' for outgoing material from the accretion disk, ``ejecta'' for ejected material from the outer surface of the WR. } \citep{OMN05a}. 
This wind, in turn, injects a considerable amount of energy into the rest of the ejecta, 
leading to a Primary-Induced Accretion Transient (PIAT). 
The schematic outline of a PIAT is shown in Figure \ref{fig:schematic}. 

On the other hand, for more massive systems of $M_*\sim10^{1.5}\rm~\msun$ and $a\sim10^{12}$ cm, 
the synchronization time is shorter than the evolution timescale of massive stars, 
($t_{\rm TL}\sim7.5\times10^{4}$~yr). Then, the spin of the secondary can be tidally synchronized 
to its orbital period \citep[cf.][]{DLP08a,YWL10a}.  In this case, the outer material of the secondary 
star has sufficient angular momentum to form a disk as it gravitationally collapses. 
The disk is massive enough to produce a powerful wind, 
which results in a Tidally-Locked Secondary Supernova (TLSSN). 
A bright radio afterglow is also expected owing to the high kinetic energy of the wind.
These TLSSNe are discussed in the accompanying paper \citep[][; Paper I]{KMM17a}.

Both PIATs and TLSSNe are caused by winds from accretion disks when BBHs are formed through binary evolution. Optical transients caused by such winds have also been considered in the context of hypernovae or neutron star mergers with radioactive nuclei \citep{MW99a,PR06a,KFM15a,KSK15a}, super-luminous supernovae \citep{DK13a}, single BH formation~\citep{KQ15a} and BH mergers~\citep{MKM16a}. 

In this paper, we focus on PIATs, in which the binaries are not tidally synchronized. 
We analytically estimate the wind luminosity in Section \ref{sec:wind}, and calculate the features 
of the resulting optical transient in Section \ref{sec:PIAT}.  
We discuss the observational prospects and caveats in Section \ref{sec:discussion}, and summarize our 
results in Section \ref{sec:summary}. 
We use the notation of $A=A_x10^x$ throughout this work.

\section{Disk-Driven Wind from the Primary Black Hole}\label{sec:wind}

We consider a massive binary system consisting of a primary BH and a secondary WR. 
The binary parameters are: the mass of the secondary WR $M_*=10\rm~\msun$, the radius of the WR $R_*=10^{11.5}$~cm 
\footnote{The radii of WR stars are somewhat uncertain. An atmospheric model suggests $R_*\sim 2\times10^{11}$ cm \citep{Cro07a}. 
Stellar evolution models predict a relation $R_*\sim 7\times10^{10}(M_*/10~\msun)^{0.7}\rm~cm$ \citep{SM92a,KZK16a}, 
while a binary evolution model shows that the lighter secondary makes the radius larger, $R_*\sim 10^{12}$ cm for $M_*\sim5\msun$ \citep{YWL10a}.}
, a separation $a=10^{12}$~cm, and a primary mass $M_{\rm BH}$. 
Hereafter, we fix $M_{\rm BH}=M_*$ for simplicity. 
The orbital velocity of the BH is $v_{\rm orb}=\sqrt{GM_*/(2a)}\simeq4.6\times10^7M_{*,1}^{1/2}a_{12}^{-1/2}\rm~cm~s^{-1}$. 
When the WR collapses to a BH, its outer material of mass $M_{\rm ej}\sim10^{-2}\msun$ may be ejected \citep{FQK17a}.
The ejecta velocity is comparable to the escape velocity, 
\begin{eqnarray}
V_{\rm ej,i}&\approx&\xi V_{\rm esc}=\xi\sqrt{\frac{2GM_*}{R_*}} \\
  &\simeq&1.8\times10^8M_{*,1}^{1/2}R_{*,11.5}^{-1/2}\xi_{0.3}\rm~cm~s^{-1},\label{eq:veji}\nonumber
\end{eqnarray}
where $\xi\sim2$ is a correction factor relative to the escape velocity. 
This ejecta velocity is always higher than the orbital velocity. 
The kinetic energy of the ejecta is estimated to be
\begin{equation}
{\mathcal E}_{\rm ej,i}= \frac{1}{2}M_{\rm ej}V_{\rm ej,i}^2
  \simeq3.3\times10^{47}M_{*,1}R_{*,11.5}^{-1} M_{\rm ej,-2}\xi_{0.3}^{2}\rm~erg,\label{eq:ekineji}
\end{equation}
This expelled ejecta amounts to an explosion energy comparable to the gravitational energy loss due to the neutrino radiation \citep{Nad80a,LW13a}.

We assume that the ejecta is in homologous expansion. 
The density profile inside the ejecta is often parameterized through a power-law form with index 
0--1 \citep[e.g.][]{LP98a,KB10a}, and here we assume a uniform density (index $=0$) for simplicity. 
The velocity profile inside the homologously expanding ejecta is written as $V(R)=R/t'$  ($t'=0$ is the time when the secondary collapses). 
During the expansion, as the velocity at fixed radius decreases with time, the gas that has a velocity 
lower than $V_{\rm esc}$ cannot escape to infinity. We can introduce a threshold velocity, 
$V_{\rm thr}=V_{\rm esc}\chi'=V_{\rm esc}\sqrt{1-R_*/a}$, such that the material with 
$V(R_*)<V_{\rm thr}$ is not ejected and falls back into the secondary. Note that for this threshold 
velocity, all the material that can reach the position of the primary is ejected. 
Then, the density of the ejecta at $R=a$ can be expresed as 
\begin{equation}
 \rho_{\rm ej,m}\approx\frac{3M_{\rm ej}}{4\pi a^3}\left(\frac{t'}{t_{\rm arr}'}\right)^{-3}\left(\Theta(V_{\rm ej,i}t'-a)-\Theta(V_{\rm thr}t'-a)\right),\label{eq:densm}
\end{equation}
where $\Theta(x)$ is the Heaviside step function and $t_{\rm arr}'=a/V_{\rm ej,i}$ is the arrival time of 
the ejecta at the primary position. The fraction of the fallback matter is small, $M_{\rm fb}/M_{\rm ej}\approx(V_{\rm thr}/V_{\rm ej,i})^3\simeq7.1\times10^{-2}\xi_{0.3}^{-3}\chi_{-0.083}'^{3}$.

After  $t'>t_{\rm arr}'$, the ejecta accretes onto the primary BH. 
Assuming that the sound speed in the ejecta is small due to adiabatic expansion,
the accretion radius is $R_{\rm acc}=GM_{\rm BH}/(V_a^2+v_{\rm orb}^2)\approx GM_{\rm BH}/V_a^2$, 
where $V_a=a/t$ is the ejecta velocity at $R=a$ \citep{Edg04a}. 
Since $R_{\rm acc}<a$ is satisfied, 
we can estimate the accretion rate to be given by the Bondi-Hoyle-Lyttleton rate \citep{SMT85a,Edg04a}
\begin{eqnarray}
 &\mdotbh&\approx4\pi R_{\rm acc}^2\rho_{\rm ej,m}\sqrt{V_a^2+v_{\rm orb}^2}\approx\frac{3M_{\rm ej}(GM_{\rm BH})^2}{a^3V_{\rm ej,i}^3} \label{eq:mdotbh} \\
&\simeq&8.7\times10^{-9}M_{*,1}^{1/2}R_{*,11.5}^{3/2}M_{\rm ej,-2}a_{12}^{-3}\xi_{0.3}^{-3} \rm~\msun~s^{-1}.\nonumber
\end{eqnarray}
For our conditions this accretion rate is approximately constant in time, 
and much higher than the Eddington accretion rate, 
$\dot M_{\rm Edd}=L_{\rm Edd}/c^2\simeq 7.0\times10^{-16}M_{*,1}\rm~\msun~s^{-1}$. 
The duration of this high accretion rate is 
\begin{equation}
 t_{\rm dur}=t_{\rm stop}'-t_{\rm arr}'\simeq7.7\times10^3M_{*,1}^{-1/2}R_{*,11.5}^{1/2}a_{12}\chi_{-0.16}\rm~s,
\end{equation}
where $t_{\rm stop}'=a/V_{\rm thr}$ and $\chi=(\xi-\chi')/(\xi\chi')\simeq0.7$ for the reference parameters. 
After this time, the accretion stops because there is no gas around the primary due to the fallback. 
The total accreted mass, 
$M_{\rm acc}\approx\mdotbh t_{\rm dur}\simeq6.7\times10^{-5}R_{*,11.5}^{2}M_{\rm ej,-2}a_{12}^{-2}\xi_{0.3}^{-3}\chi_{-0.16}\rm~\msun$, 
is much smaller than $M_{\rm ej}$. 

The accreted gas, due to its orbital angular momentum, forms a disk surrounding the primary BH \citep{VKS09a,HCN13a}, 
The high accretion rate leads the disk to the advection dominated regime \citep{abr+88}, where the wind production is expected \citep{ny94,BB99a}. Numerical simulations suggest the existence of powerful disk winds for ultraluminous X-ray sources via radiation \citep{OMN05a,JSD14a} and gamma-ray bursts via viscous heating and/or magnetohydrodynamic turbulence \citep{MW99a,DOB09a,FM13a,KSK15a}. 
We assume that the wind is almost isotropic, a fraction $\eta_w\sim1/3$ of the accreted material going into the wind, 
whose velocity is approximately constant, $V_w\sim10^{10}\rm~cm~s^{-1}$. 
These values of $\eta_w$ and $V_w$ are consistent with recent radiation magnetohydrodynamic calculations \citep{TO15a,NSS17a} and observations of ultrafast outflows in quasars \citep{HOD15a},
although these values have large uncertainty.
The luminosity and the total energy of the wind are 
\begin{eqnarray}
 L_w&\approx&\frac{1}{2}\eta_w\mdotbh V_w^2\simeq2.7\times10^{44}M_{*,1}^{1/2}R_{*,11.5}^{3/2}\nonumber  \\
& \times& M_{\rm ej,-2}a_{12}^{-3}\xi_{0.3}^{-3}\eta_{-0.5}V_{10}^2  \rm~erg~s^{-1},\label{eq:lw_comp}\\
{\mathcal E}_w&\approx& L_wt_{\rm dur}\simeq2.1\times10^{48}R_{*,11.5}^{2}M_{\rm ej,-2}\nonumber\\
&\times&a_{12}^{-2}\xi_{0.3}^{-3}\chi_{-0.16}\eta_{-0.5}V_{10}^2 \rm~erg.
\end{eqnarray}
Note that although the opening angle of the wind is large, 
it is not isotropic and probably does not cover the entire solid angle (see panel (b) of Figure \ref{fig:schematic}). 
We assume that the mass accretion onto the BH is not quenched by the wind. 

As the material accretes onto the BH through an accretion column behind the BH,
the resulting accretion disk is initially covered by the ejecta, before producing the wind (see the panel (a) of Figure \ref{fig:schematic}). 
Thus, the wind is, at least initially, confined within the ejecta.
When the wind collides with the ejecta, the forward shock and reverse shock propagate in the ejecta and the wind, respectively. All the kinetic energy of the wind dissipated in the reverse shock is converted into radiation energy efficiently (see Subsection \ref{sec:caveats}). Since the photons produced are trapped in the ejecta, they can accelerate or heat up the ejecta.

\section{Primary-Induced-Accretion Transients from newborn BBHs (BBH-PIATs)}\label{sec:PIAT}

\subsection{PIAT-I: The cases with $\mathcal{E}_w>\mathcal{E}_{\rm ej,i}$}\label{sec:bright}

As a fiducial case, we consider the cases with $\mathcal{E}_w>\mathcal{E}_{\rm ej,i}$,
where the radiation energy injected by the wind accelerates the ejecta in a sound crossing time. We approximately consider the instant acceleration of the ejecta, since sound crossing time is shorter than the photon diffusion time. 
For $t<t_{\rm dur}$ ($t=t'-t_{\rm arr}'$), during which the wind is produced,
the ejecta velocity is estimated to be $V_{\rm ej}\approx\sqrt{L_wt/M_{\rm ej}}\propto t^{1/2}$, and the ejecta radius is $R_{\rm ej}\approx V_{\rm ej}t\propto t^{3/2}$. 
The total internal energy inside the ejecta is determined by the balance between adiabatic losses and energy injection, $\mathcal{E}_{\rm int}/t_{\rm dyn}\sim L_w$, where $t_{\rm dyn}=R_{\rm ej}/V_{\rm ej}\approx t$. 
This leads to $\mathcal{E}_{\rm int}\approx L_w t \propto t$.
Although the ejecta confines the bulk of the photons, 
a small fraction of photons can diffuse out from the surface of the ejecta. 
This photon diffusion luminosity evolves as $L_{\rm ph}\sim {\mathcal E}_{\rm int}/t_{\rm ph}\propto t^{5/2}$, 
where $t_{\rm ph}\approx R_{\rm ej}\tau/c\sim M_{\rm ej}\kappa/(cR_{\rm ej})$ ($\tau\approx\rho_{\rm ej}\kappa R_{\rm ej}$ and $\kappa$ is the opacity\footnote{Note that the ejecta density $\rho_{\rm ej}$ is different from the $\rho_{\rm ej,m}$ representing the ejecta density that is not accelerated by the wind.}). 
The effective temperature is estimated to be $T_{\rm eff}=(L_{\rm ph}/(4\pi\sigma R_{\rm ej}^2))^{1/4}\propto t^{-1/8}$, where $\sigma$ is the Stefan-Boltzmann constant.

\begin{figure}[tbp]
   \begin{center}
   \includegraphics[width=\linewidth]{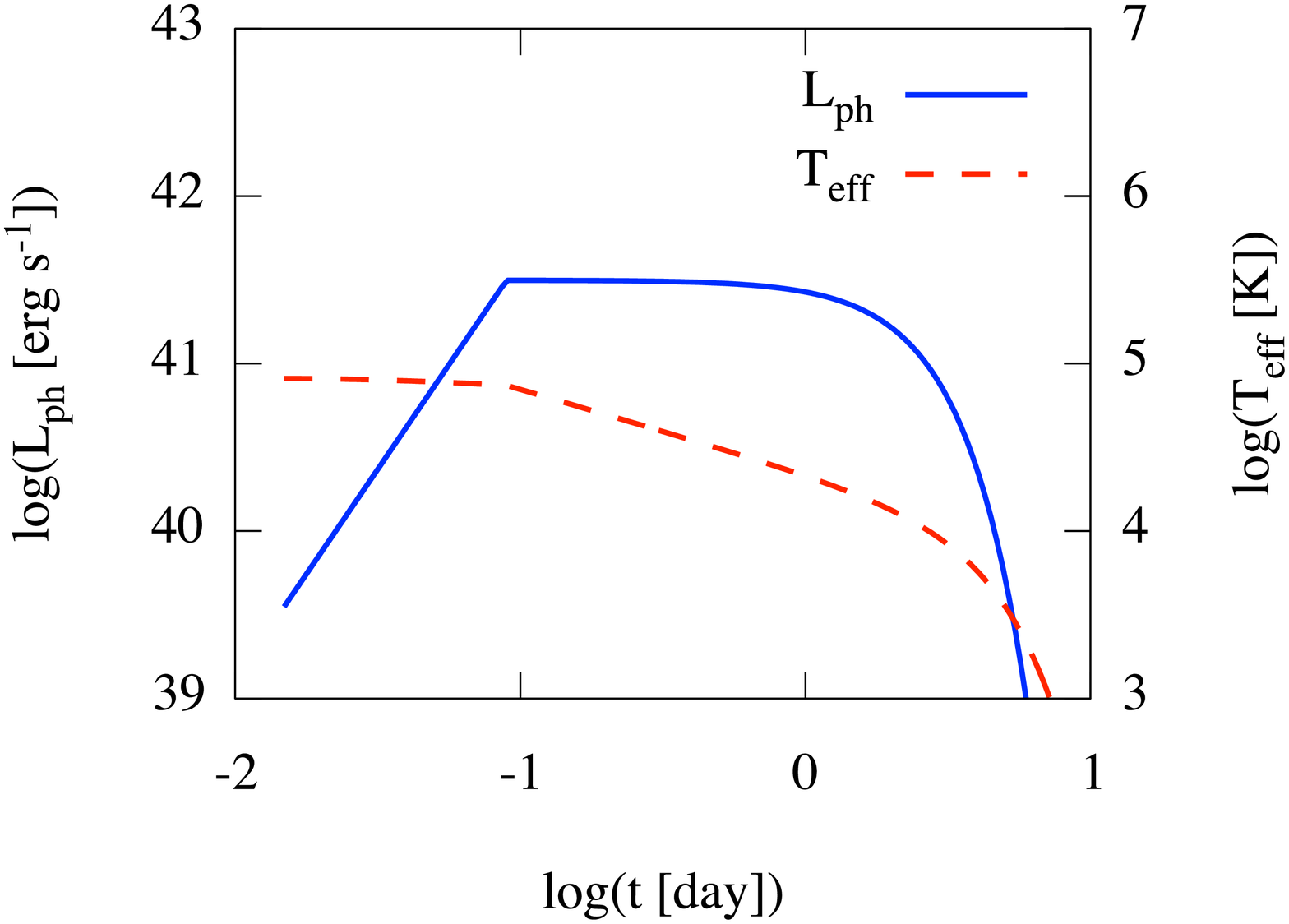}
   \includegraphics[width=\linewidth]{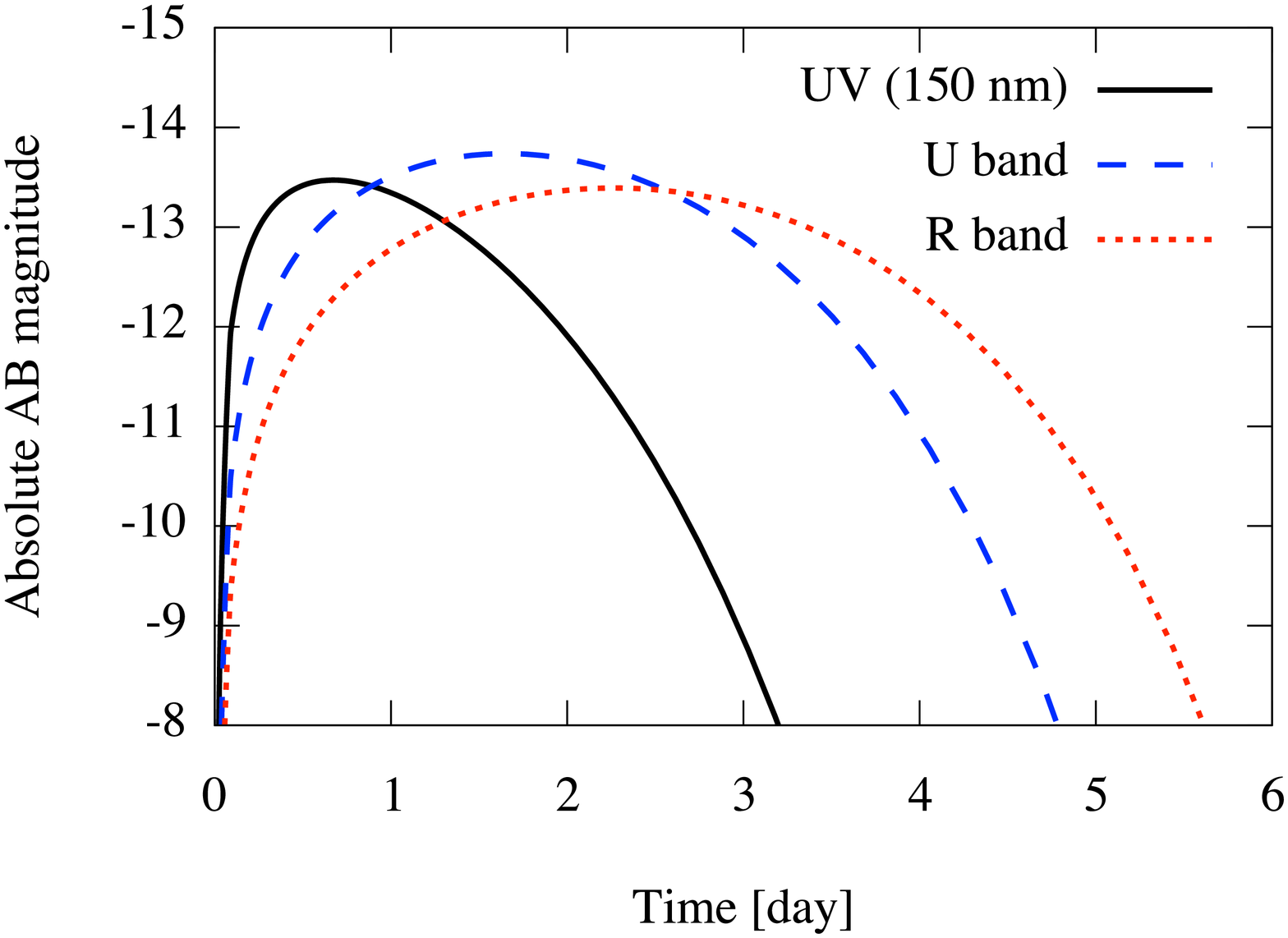}
    \caption{Upper panel: time evolution of the bolometric luminosity $L_{\rm ph}$ (blue-solid) and the effective temperature $T_{\rm eff}$ (red-dashed) for PIAT-I. Lower panel: time evolution of absolute AB magnitudes for the UV range (150 nm, black-solid), $U$ band (blue-dashed), and $R$ band (red-dotted) for PIAT-I. The parameters are $M_*=10\rm~M_{\odot}$, $R_*=10^{11.5}$ cm, $a=10^{12}$ cm, $\xi=10^{0.3}$, $\eta_w=10^{-0.5}$, and $V_w=10^{10}\rm~cm~s^{-1}$.}
    \label{fig:evolv}
   \end{center}
  \end{figure}

When the wind stops at $t=t_{\rm dur}$, the ejecta velocity is estimated to be $V_{\rm ej,c}\approx\sqrt{2{\mathcal E}_w/M_{\rm ej}}$, and the internal energy is $\mathcal{E}_{\rm int}\approx\mathcal{E}_w$. 
For $t>t_{\rm dur}$, the ejecta velocity is approximately constant $V_{\rm ej}=V_{\rm ej,c}$,
and the radius evolves as $R_{\rm ej}\approx V_{\rm ej,c}t$. 
The adiabatic expansion determines the evolution of the internal energy, $d\mathcal{E}_{\rm int}/dt = \mathcal{E}_{\rm int}/t_{\rm dyn}$.
This leads to $\mathcal{E}_{\rm int}\approx \mathcal{E}_wt_{\rm dur}/t$.
In this phase, there is a cavity inside the ejecta,
so the density is written as $\rho_{\rm ej}\approx M_{\rm ej}/(4\pi R^2H)\propto t^{-3}$,
where $H\propto R_{\rm ej}$ is the thickness of the ejecta.
The photon diffusion time is estimated to be $t_{\rm ph}\approx R_{\rm ej}H\rho_{\rm ej}\kappa/c\propto t^{-1}$.
The luminosity of the diffusing photons is then constant in time, $L_{\rm ph}\sim \mathcal{E}_{\rm int}/t_{\rm ph}\propto t^0$. 
The effective temperature evolves as $T_{\rm eff}\propto (L_{\rm ph}/R_{\rm ej}^2)^{1/4} \propto t^{-1/2}$.
This phase continues until the radiation breakout time at which the bulk of the photons escape from the ejecta,
\begin{eqnarray}
 t_{\rm bo}&\approx&\sqrt{\frac{\kappa M_{\rm ej}}{4\pi cV_{\rm ej,c}}}\simeq1.5\times10^{5}R_{*,11.5}^{-1/2}\nonumber\\
&\times&M_{\rm ej,-2}^{1/2}a_{12}^{1/2}\xi_{0.3}^{3/4}\chi_{-0.16}^{-1/4}\eta_{-0.5}^{-1/4}V_{10}^{-1/2}\rm~s,
\end{eqnarray}
where we use $\kappa\simeq 0.2\rm~cm^2~g^{-1}$ that corresponds to fully ionized helium. 
The internal energy at the radiation breakout time is $\mathcal{E}_{\rm bo}\approx \mathcal{E}_wt_{\rm dur}/t_{\rm bo}$.
The luminosity and effective temperature at the radiation breakout time are then computed as
\begin{eqnarray}
 L_{\rm bo}&\approx&\frac{\mathcal{E}_{\rm bo}}{t_{\rm bo}}\simeq7.1\times10^{41}M_{*,1}^{-1/2}R_{*,11.5}^{7/2}\nonumber\\
  &\times&a_{12}^{-2}\xi_{0.3}^{-9/2}\chi_{-0.16}^{5/2}\eta_{-0.5}^{3/2}V_{10}^{3}\rm~erg~s^{-1},\label{eq:lph}\\
 T_{\rm eff,bo}&=&\frac{L_{\rm bo}}{4\pi\sigma R_{\rm bo}^2}
  \simeq2.1\times10^4 M_{*,1}^{-1/8}R_{*,11.5}^{5/8}M_{\rm ej,-2}^{-1/4}\nonumber\\
  &\times&a_{12}^{-1/4}\xi_{0.3}^{-3/4}\chi_{-0.16}^{1/2}\eta_{-0.5}^{1/4}V_{10}^{1/2}\rm~K,\label{eq:Teff}
\end{eqnarray}
where $R_{\rm bo}=V_{\rm ej,c}t_{\rm bo}$.
For $t>t_{\rm bo}$, the photon diffusion determines the internal energy, $d{\mathcal E}_{\rm int}/dt\sim {\mathcal E}_{\rm int}/t_{\rm ph}$. 
Then, the internal energy decays as $\exp(-t^2/(2t_{\rm bo}^2))$, so both $L_{\rm ph}$ and $T_{\rm eff}$ decreases rapidly. 
We show the evolution of $L_{\rm ph}$ and $T_{\rm eff}$ in the upper panel of Figure \ref{fig:evolv}. 

We plot the UV (150 nm), $U$ band (365 nm), and $R$ band (658 nm) absolute AB magnitudes in the lower panel. Since the optical depth $\tau>1$ is satisfied for the times of interest, we use the Planck spectra corresponding to $T_{\rm eff}$. 
The peak magnitude and duration are around $-$14 and 3 days for $U$ band, respectively, with longer duration for the longer wavelength.  
These transients may be distinguishied from usual SNe by means of their bluer color and shorter duration. 
They can also be distinguished from macronovae/kilonovae because they show helium lines, as is the case 
also for TLSSNe (see Paper I). 
Note that the recombination of helium takes place at $T<10^4\rm~K$, which may affect the light curves 
for later time of $t\gtrsim3$ days. Note also that PIATs are dimmer and of longer duration, compared 
to the TLSSNe discussed in Paper I.

\subsection{PIAT-II: The case with $\mathcal{E}_w<\mathcal{E}_{\rm ej,i}$}\label{sec:PIAT-II}

For the cases with  $\mathcal{E}_w<\mathcal{E}_{\rm ej,i}$, for example $a=10^{12.5}$ cm and with the other parameters the same as the fiducial values, the ejecta is not accelerated by the wind. 
However, even in this case, the ejecta acquires a significant amount of internal energy from the wind, 
resulting in a somewhat brighter transient than that without winds (see Section \ref{sec:failedSNe}).
The model with lower value of $\eta_w$ and/or $V_w$ also leads to this regime.

The evolutionary features of the transient are similar to those in Section \ref{sec:bright} except for $V_{\rm ej}=V_{\rm ej,i}$ for the entire evolution. For $t<t_{\rm dur}$,   $R_{\rm ej}\approx V_{\rm ej,i}t\propto t$, and the internal energy is $\mathcal{E}_{\rm int}\approx L_w t\propto t$. The photon diffusion time is $t_{\rm ph}\propto t^{-1}$, so that $L_{\rm ph}\propto t^{2}$ and $T_{\rm eff}\propto t^{0}$. For $t_{\rm dur}<t<t_{\rm bo}$, the time evolutions of the physical quantities are the same as those in Section \ref{sec:bright}: $L_{\rm ph}\propto t^0$ and $T_{\rm eff}\propto t^{-1/2}$. The photon breakout time is estimated to be
\begin{equation}
 t_{\rm bo}\approx\sqrt{\frac{\kappa M_{\rm ej}}{4\pi cV_{\rm ej,i}}}\simeq2.4\times10^{5}R_{*,11.5}^{1/4}M_{*,1}^{-1/4}M_{\rm ej,-2}^{1/2}\xi_{0.3}^{-1/2}\rm~s.\label{eq:tbo_dim}
\end{equation}
The luminosity and effective temperature at the breakout time are
\begin{equation}
 L_{\rm bo}\simeq5.4\times10^{40}R_{*,11.5}^{2}a_{12.5}^{-1}\xi_{0.3}^{-2}\chi_{-0.26}^{2}\eta_{-0.5}V_{10}^{2}\rm~erg~s^{-1},\label{eq:lph-II} 
\end{equation}
\begin{eqnarray}
 T_{\rm eff,bo}\simeq1.4\times10^4 M_{*,1}^{-1/8}R_{*,11.5}^{5/8}M_{\rm ej,-2}^{-1/4}a_{12.5}^{-1/4}\nonumber\\
\xi_{0.3}^{-3/4}\chi_{-0.16}^{1/2}\eta_{-0.5}^{1/4}V_{10}^{1/2}\rm~K.\label{eq:Teff-II} 
\end{eqnarray}

  \begin{figure}[tbp]
   \begin{center}
   \includegraphics[width=\linewidth]{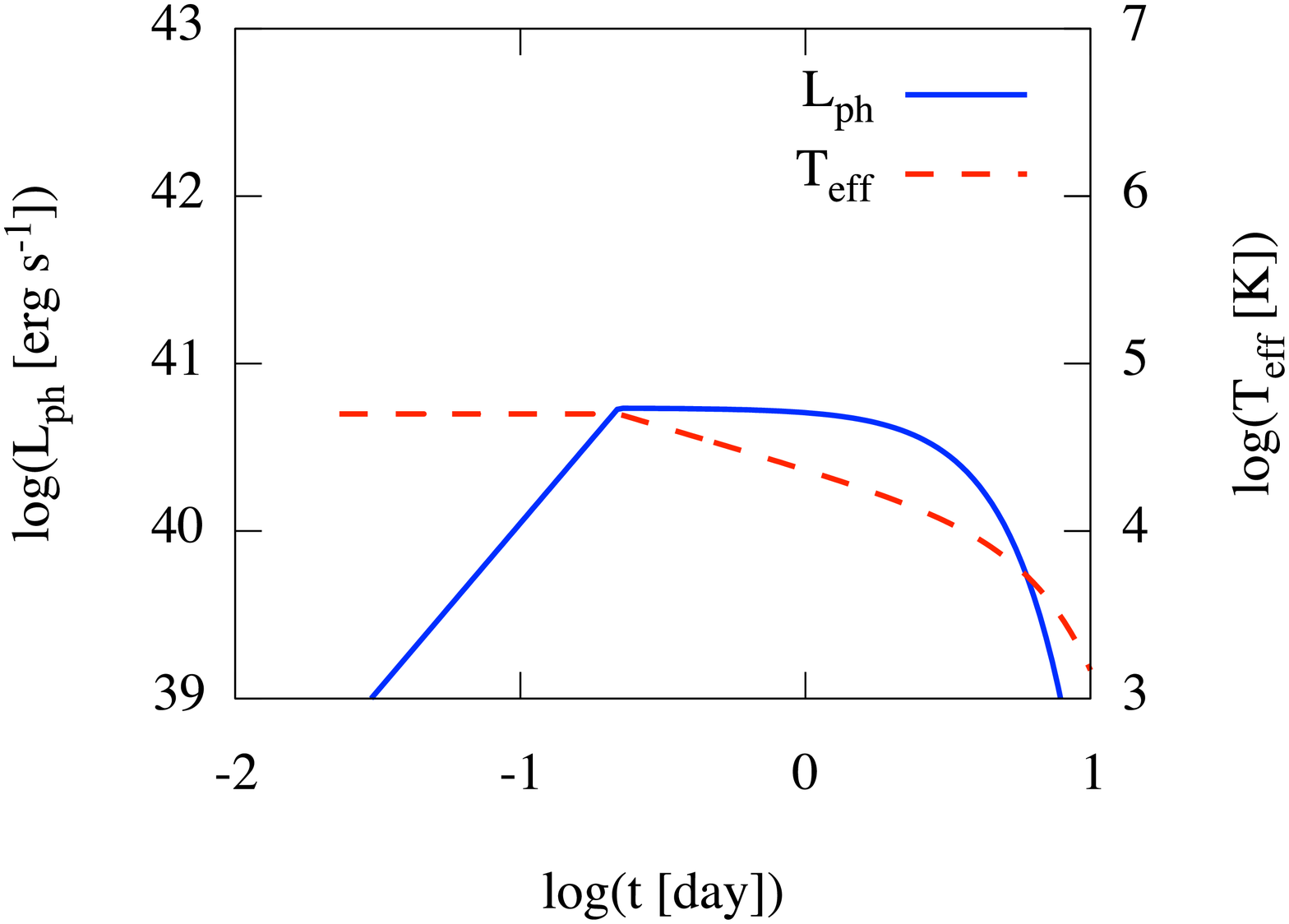}
   \includegraphics[width=\linewidth]{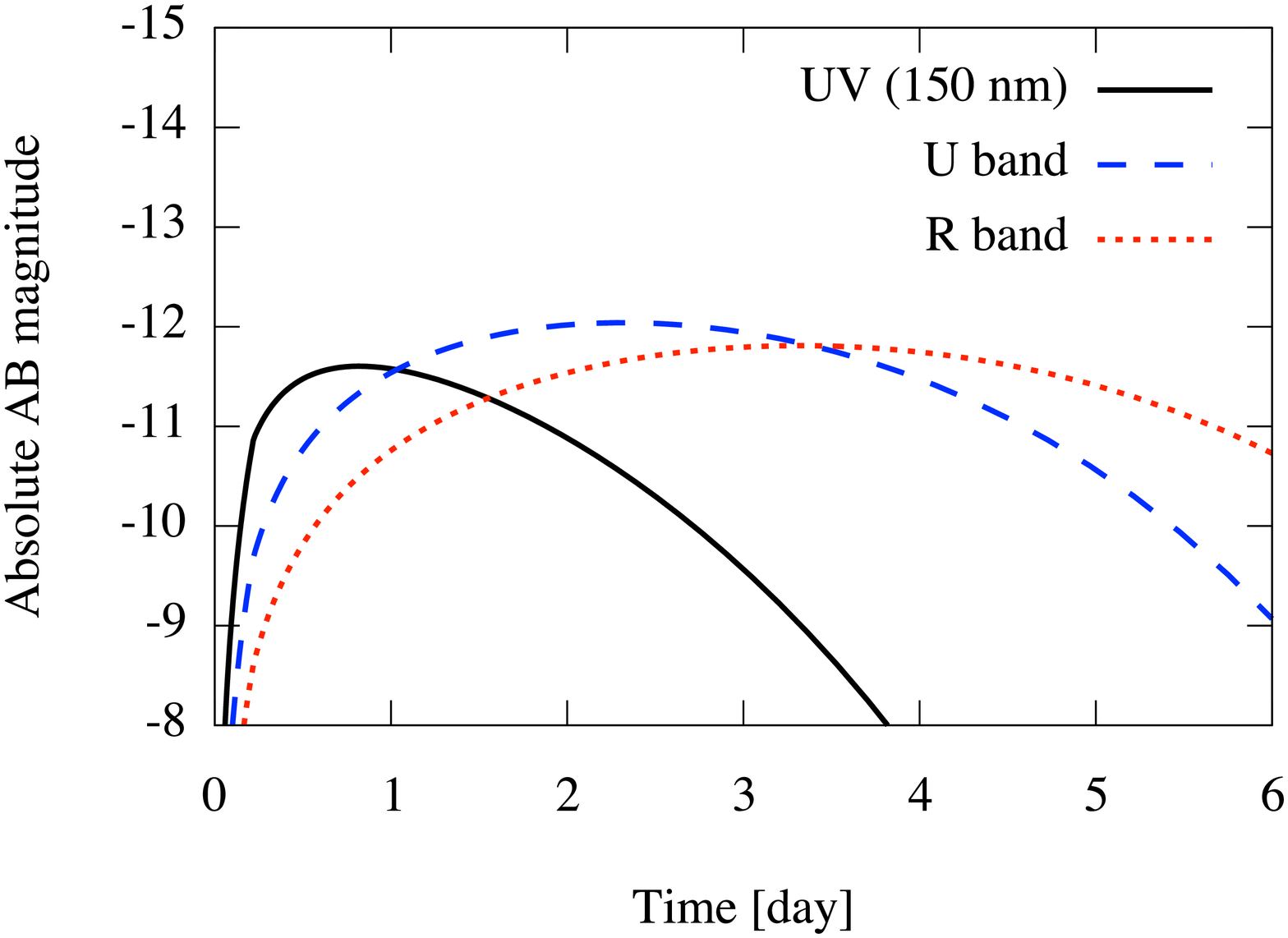}
    \caption{Same as Figure \ref{fig:evolv}, but for the case with PIAT-II ($a=10^{12.5}\rm~cm$).}
    \label{fig:dim_evolv}
   \end{center}
  \end{figure}

  Figure \ref{fig:dim_evolv} shows the time evolutions of $L_{\rm ph}$, $T_{\rm eff}$ (upper panel), and the absolute AB magnitudes for UV (150 nm), $U$ band, and $R$ band (lower panel). This model is roughly an order of magnitude dimmer than the previous PIAT-I, although the features of the light curves are similar. Also, the effect of helium recombination would be important for the late times of $t\gtrsim3$ days, as is for the PIAT-I.

\subsection{Failed SNe}\label{sec:failedSNe}

In the cases where ${\mathcal E}_w\ll{\mathcal E}_{\rm ej,i}$, taking for example a separation $a\gtrsim10^{13}$ cm and/or the wind parameters $\eta_w\lesssim0.1$ and $V_w\lesssim0.1c$, with the other parameters the same as the reference values, the energy injection by the wind is less important.
Instead a weak explosion with the initial ejecta energy of Equation (\ref{eq:ekineji}) may occur. 
We assume the initial internal energy of the ejecta is similar to ${\mathcal E}_{\rm ej,i}$. 
The radiation breakout time is the same as Equation (\ref{eq:tbo_dim}).
Considering an adiabatic evolution, $\mathcal{E}_{\rm int}\propto t^{-1}\propto R_{\rm ej}^{-1}$, 
the internal energy at the breakout time is $\mathcal{E}_{\rm bo}\approx \mathcal{E}_{\rm ej,i}R_*/(V_{\rm ej,i}t_{\rm bo})$.
The luminosity and the effective temperature are then
\begin{equation}
 L_{\rm bo}\approx\frac{\mathcal{E}_{\rm bo}}{t_{\rm bo}}\simeq1.0\times10^{40}M_{*,1}\xi_{0.3}^2\rm~erg~s^{-1}.
\end{equation}
\begin{equation}
 T_{\rm eff,bo}\simeq 9.2\times10^{3}M_{*,1}^{1/8}R_{*,11.5}^{1/8}M_{\rm ej,-2}^{-1/4}\xi_{0.3}^{3/8}.
\end{equation}
This event is dimmer than the PIATs considered above.
Such failed SNe could  however be more luminous if some amount of radioactive nuclei are produced or are included inside the ejecta \citep{MTT10a}.

\section{Discussion}\label{sec:discussion}
\subsection{Observational prospects}

For the nominal parameters adopted, one may expect an event rate for these transients similar to 
the LIGO event rate, $\sim10-200$ Gpc$^{-3}$ yr$^{-1}$ \citep{LIGO16e,LIGO17a}. 
Using the PIATs' absolute magnitude of -14, the distance at which they would be detectable with current 
surveys with sensitivities of $\sim21$ mag is around 100~Mpc, and the
event rate within this distance is 0.04--0.8 yr$^{-1}$.
Thus, if the event rate for PIATs is in the higher range of the expected values,
it is possible to detect PIATs with current surveys, such as the Panoramic Survey Telescope \& Rapid Response System \citep[Pan-STARRS;][]{Pan-STARRS04a}, the Palomar Transient Factory \citep[PTF;][]{PTF09a}, and the Kiso Supernova Survey \citep[KISS;][]{KISS14a}.
For future projects with sensitivity of $\sim25$ mag,  e.g., the Large Synoptic Survey Telescope \citep[LSST,][]{LSST09a}, the detectability distance is around 600~Mpc, and  the
expected event rate is 9--180 yr$^{-1}$, making the detection of PIATs possible. 
However, we should note that the event rate of PIATs involves some uncertainties.  
The effect of wide binaries may not be ignorable,  since the separation distribution is  flat 
in log-space of binary separation, $\propto a^{-1}$ \citep{Abt83a}.
Although PIATs in the wide binaries with WRs are faint, 
the wide binaries with different progenitors (BSGs or RSGs) may increase the event rate. 
Also, the star formation rate is higher for higher redshift, 
so we could expect that the BBH formation rate has the same tendency \citep[e.g.,][]{IKV16a,MM16a}.
In this case, the event rate in the local universe is likely to be lower than the GW rate.

This type of transient can also be observed in soft X-ray surveys, 
such as the extended ROentgen Survey with an Imaging Telescope Array \citep[eROSITA;][]{eROSITA12a}.
If the total wind energy is higher than the initial ejecta energy, a forward shock produced by the wind can break out from the ejecta \citep{MM99a,MVH14a,MKK15a}.
A detailed prediction would require at least one-dimensional radiation hydrodynamical modeling, which is beyond the scope of this paper.

There could be other PIATs occurring throughout the life of a BH-massive star binary. One possibility 
arises during the common envelope phase, in which the secondary ejects a large amount of  hydrogen 
envelope \citep{BHB16a}. 
The primary BH accretes the common envelope of $\rho\sim3M_{\rm env}/(4\pi a^3)\sim1\times10^{-8}\rm~g~cm^{-3}$ for $M_{\rm env}\sim30\rm~\msun$ and $a\sim10^{14}$ cm. 
Using the Bondi-Hoyle-Lyttleton rate, the mass accretion rate is estimated to be $2\times10^{-6}\rm~\msun~s^{-1}$, 
where we use $v\sim\sqrt{GM_{\rm BH}/a}\sim4\times10^6\rm~cm~s^{-1}$. 
Thus, it might be possible to have luminous transients powered by this huge accretion luminosity. 
Another possibility are supernova impostors, in which the evolved secondary ejects a large amount of its envelope \citep{SLS11a}.  The mass of the ejected material is estimated to be 0.01--10 $\rm\msun$. 
Since the accretion rate and the wind luminosity are proportional to the ejecta mass, the supernova 
impostors could in principle induce more luminous PIATs than what occurs at the BBH formation.

The PIAT transients considered here leave an accretion disk around a BH in a BBH, which a few years later becomes neutral due to radiative cooling, thus  naturally creating a fossil disk in which the angular momentum transport is inefficient \citep[e.g.,][]{PDC14a,PLG16a,KTT17a}. 
This fossil disk can remain for millions of years until the BBH merges, resulting in the possible electromagnetic counterparts of GWs from BBHs \citep{MKM16a,KTT17a,MK17a}.

The spin of the primary BH can also affect the light curves of PIATs. If the primary has a high spin, 
a relativistic jet is launched when the primary accretes the ejecta \citep{BZ77a,Kom04a,TT16a}. 
If the jet can penetrate the ejecta, GRBs of long duration ($t\sim 10^3$ s) and very low-luminosity ($L_{\rm iso}\sim 10^{47}\rm~erg~s^{-1}$) may be possible.
On the other hand, if the jet dissipates most of its kinetic energy inside the ejecta, the resultant transients might be similar to PIATs.
The condition for the jet to penetrate the ejecta depends on the detail of the geometry, which is beyond the scope of this paper.

Other formation channels to form a BBH involve binaries consisting of a BH and a blue supergiant (BSG) \citep{KIH14a,IHK17a}.  
Since some BSGs have very compact cores, they can eject very little amounts of their outer material, which is unlikely to produce PIATs. If the cores of BSGs are not very compact, BSGs can eject their envelopes of $\sim 0.1\rm~M_{\odot}$ \citep{FQK17a}.
Red supergiants (RSGs), which also collapse to BHs, can eject large amount of their envelopes ($\gtrsim1\rm~\msun$) when collapsing \citep{LW13a}.
Thus, it is possible that BH-RSG and BH-BSG binaries could produce PIATs.  
However, our spherically symmetric treatment would not be accurate for such cases, because of the very large separation ($a>R_*\gtrsim10^{13}$ cm).

\subsection{Caveats}\label{sec:caveats}

Although for simplicity we have used a spherically symmetric formulation, there are substantial 
non-spherical effects possible in this system. 
One is the effect of the finite binary separation, which means that the energy source (primary BH) is not located at the center of the ejecta. 
This might affect the initial evolution of the ejecta, whose quantitative discussion would probably be subject to change. 
Another is the effect of the wind.  We assumed some tuning so that the opening angle is wide enough 
that we can approximate it as spherically symmetric, while the wind is not completely isotropic so 
that the ejecta density profile at the mid-plane does not change much. 
In reality, the wind from an accretion flow might be more bipolar-like \citep{SNM14a,TOK16a}. 
The ejecta density profile is also probably affected by the wind, so that the accretion rate might 
be modified.  It is beyond the scope of this paper to investigate these effects, 
which would require 3-dimensional radiation hydrodynamic calculations. 

We assumed for simplicity that all the kinetic energy of the wind is converted into the radiation 
energy of the ejecta, and we now discuss the validity of this assumption. 
The wind kinetic energy is first converted to the thermal energy of protons at the reverse shock. 
The electrons inside the shocked region are heated up by Coulomb collisions and other plasma processes. 
When the electrons become energetic enough, 
the bremsstrahlung cooling time becomes shorter than the Coulomb loss timescale \citep{tk85}. 
In addition, the Compton cooling is also efficient for the relativistic electrons.  
If we consider only the Coulomb heating in the far downstream, 
the electron temperature may be around $k_{\rm B}T_e\sim m_e c^2$. 
Then, the energy loss time of the protons due to Coulomb losses can be estimated to be 
$t_{\rm cool}\approx(m_p/m_e)/(\rho_{\rm ps} \kappa c \ln \Lambda)\simeq1.3\times10^3 M_{*,1}^{-3/2}R_{*,11.5}^{3/2}a_{12}^{3}M_{\rm ej,-2}\chi_{-0.16}^{3}V_{10}^{3} \rm~s$, 
where $\rho_{\rm ps}$ is the density in the post shock region and $\ln \Lambda \sim30$ is the Coulomb logarithm  \citep{tk85,ktt14}. 
Here, setting $t=t_{\rm dur}$, we have used $\rho_{\rm ps}\sim L_w/(2\pi  R_{\rm ws}^2V_w^3)$. 
Since $t_{\rm cool}<t_{\rm dyn}\approx t_{\rm dur}$ is satisfied at $t=t_{\rm dur}$, 
the wind kinetic energy can be converted into radiation energy. 
In reality, the Compton cooling can also be relevant, 
so that the cooling of electrons may be even stronger than the above estimate.

The parameters for the wind are also subject to large uncertainties. 
We implicitly assumed that the circularization radius of the accreted material is not large. 
However, if the circularization radius is too large, 
a considerable amount of gas could escape from the accretion disk as an wind before arriving near the primary BH \citep{HOK15a}. 
This  may reduce the net accretion rate, causing the PIATs to be fainter. 
Besides, although we use parameter sets consistent with recent simulations and observations 
\citep{HOD15a,TO15a,NSS17a}, lower $V_w$ and $\eta_w$ are possible. In this case, PIATs would be dimmer 
and redder as shown in Equations (\ref{eq:lph}), (\ref{eq:Teff}), (\ref{eq:lph-II}), 
and (\ref{eq:Teff-II}). 

Another caveat is that we used a one-zone approximation to estimate the light curves. 
In reality, the physical quantities have a radial dependence which may complicate the features of 
the transients. 
The circum-binary medium is possibly polluted by material ejected by a stellar wind, ejection of the common envelope and/or supernova impostors, which could affect the light curves. 
Atomic recombination processes may also modify the light curves when the ejecta cools down to a 
few $10^4$ K. One dimensional modeling including above effects gives us more precise predictions.

Finally, we mention the parameter dependence of PIATs.
For bright PIAT-Is to occur, ${\mathcal E}_w>{\mathcal E}_{\rm ej,i}$ should be satisfied. 
This condition breaks down for models with a few times smaller $R_*$, 
larger $a$, higher $\xi$, or lower $V_w$ than the reference model. 
In these cases, we expect the dimmer transients (PIAT-IIs or failed SNe; see Section \ref{sec:PIAT-II} and \ref{sec:failedSNe}). 
Considering models with higher ${\mathcal E}_w$ would be a bit extreme, 
since the reference model is fairly optimistic. 
Besides, $M_*$ should be around 10 $\msun$ because tidal synchronization might take place for 
higher $M_*$ (see Paper I), and a BH is not formed for lower $M_*$. 
Models with lower $\eta_w$ and higher or lower $M_{\rm ej}$ are also possible, which would change 
the luminosity as shown in Equations (\ref{eq:lph}) and (\ref{eq:lph-II}). 
These parameters have their distributions, and the luminosity function could be calculated using
these parameter distributions.  However, most of the parameters related to binaries, ejecta, and 
disk winds are quite uncertain.  We need to understand better the winds from super-Eddington accretion 
and the evolution of massive star binaries, in order to obtain a reliable luminosity function.

\section{Summary}\label{sec:summary}

We investigated a type of transients named PIATs, which arise from newborn BBHs formed from BH-WR 
binaries, within the context of isolated binary evolution scenarios. 
When the secondary collapses to a BH, it ejects a fraction of the outer material of the secondary. Then, a part of the ejecta naturally accretes onto the primary BH, and the accretion rate can exceed the Eddington rate, owing to the high density of the ejecta. 
As a result, the primary BH may produce a wind which injects a significant amount of energy into the 
non-accreted ejecta.  
This powers a PIAT whose kinetic energy reaches $\sim10^{48}$ ergs for optimistic  parameters.  The bolometric luminosity of this transients can be $\sim10^{40}-10^{42}\rm~erg~s^{-1}$, and the $U$-band absolute magnitude ranges from $\sim-11$ to $\sim-14$, with a duration of around a few days. 

PIATs can be distinguished from usual SNe by their weaker peak luminosity and shorter duration, and from macronovae/kilonovae by their strong helium lines. 
Recently, rapid transients of timescale around a day to a week have been observed \citep{DCS14a,TTM16b}.
Although the timescale of the observed transients is comparable to our predictions, the observed magnitudes seem much smaller. 
 
Since the systems considered in this work are extremely close binaries, the tidal force could distort the structure of the WRs, which could be considerably different from that predicted by the spherically symmetric stellar model. However, modeling the binary evolution including tidal effects is beyond the scope of this paper, which remains as a future work. 
The tidal effect is also expected to synchronize the spin period of the WR to its orbital period. In the accompanying paper \citep{KMM17a}, we propose wind-driven transients associated with the BBH formation, in which the tidal synchronization takes place in the secondary star.

\acknowledgments
The authors thank Kazumi Kashiyama, Kunihito Ioka, and Tomoya Kinugawa for useful comments.  This work is partially supported by Alfred P. Sloan Foundation (K.M.), NSF Grant No. PHY-1620777 (K.M.), NASA NNX13AH50G (S.S.K. and P.M.), an IGC post-doctoral fellowship program (S.S.K).





\begin{thebibliography}{70}
\expandafter\ifx\csname natexlab\endcsname\relax\def\natexlab#1{#1}\fi

\bibitem[{{Abbott} {et~al.}(2016{\natexlab{a}}){Abbott}, {Abbott}, {Abbott},
  {Abernathy}, {Acernese}, {Ackley}, {Adams}, {Adams}, {Addesso}, {Adhikari},
  \& et~al.}]{LIGO16e}
{Abbott}, B.~P. {et al.}\  2016{\natexlab{a}}, Physical Review X, 6, 041015

\bibitem[{{Abbott} {et~al.}(2016{\natexlab{b}}){Abbott}, {Abbott}, {Abbott},
  {Abernathy}, {Acernese}, {Ackley}, {Adams}, {Adams}, {Addesso}, {Adhikari},
  \& et~al.}]{LIGO16a}
--- 2016{\natexlab{b}}, Physical Review Letters, 116, 061102

\bibitem[{{Abbott} {et~al.}(2017){Abbott}, {Abbott}, {Abbott}, {Acernese},
  {Ackley}, {Adams}, {Adams}, {Addesso}, {Adhikari}, {Adya}, \&
  et~al.}]{LIGO17a}
--- 2017, Physical Review Letters, 118, 221101

\bibitem[{{Abramowicz} {et~al.}(1988){Abramowicz}, {Czerny}, {Lasota}, \&
  {Szuszkiewicz}}]{abr+88}
{Abramowicz}, M.~A., {Czerny}, B., {Lasota}, J.~P., \& {Szuszkiewicz}, E. 1988,
  \apj, 332, 646

\bibitem[{{Abt}(1983)}]{Abt83a}
{Abt}, H.~A. 1983, \araa, 21, 343

\bibitem[{{Belczynski} {et~al.}(2016){Belczynski}, {Holz}, {Bulik}, \&
  {O'Shaughnessy}}]{BHB16a}
{Belczynski}, K., {Holz}, D.~E., {Bulik}, T., \& {O'Shaughnessy}, R. 2016,
  \nat, 534, 512

\bibitem[{{Blandford} \& {Begelman}(1999)}]{BB99a}
{Blandford}, R.~D. \& {Begelman}, M.~C. 1999, \mnras, 303, L1

\bibitem[{{Blandford} \& {Znajek}(1977)}]{BZ77a}
{Blandford}, R.~D. \& {Znajek}, R.~L. 1977, \mnras, 179, 433

\bibitem[{{Bondi}(1952)}]{Bon52a}
{Bondi}, H. 1952, \mnras, 112, 195

\bibitem[{{Crowther}(2007)}]{Cro07a}
{Crowther}, P.~A. 2007, \araa, 45, 177

\bibitem[{{de Mink} \& {King}(2017)}]{MK17a}
{de Mink}, S.~E. \& {King}, A. 2017, \apjl, 839, L7

\bibitem[{{de Val-Borro} {et~al.}(2009){de Val-Borro}, {Karovska}, \&
  {Sasselov}}]{VKS09a}
{de Val-Borro}, M., {Karovska}, M., \& {Sasselov}, D. 2009, \apj, 700, 1148

\bibitem[{{Dessart} {et~al.}(2009){Dessart}, {Ott}, {Burrows}, \& {Rosswog}}]{DOB09a}
{Dessart}, L., {Ott}, C.~D., {Burrows}, A., {Rosswog}, S. 2008,
  \apj, 690, 1681

\bibitem[{{Detmers} {et~al.}(2008){Detmers}, {Langer}, {Podsiadlowski}, \&
  {Izzard}}]{DLP08a}
{Detmers}, R.~G., {Langer}, N., {Podsiadlowski}, P., \& {Izzard}, R.~G. 2008,
  \aap, 484, 831

\bibitem[{{Dexter} \& {Kasen}(2013)}]{DK13a}
{Dexter}, J. \& {Kasen}, D. 2013, \apj, 772, 30

\bibitem[{{Drout} {et~al.}(2014){Drout}, {Chornock}, {Soderberg}, {Sanders},
  {McKinnon}, {Rest}, {Foley}, {Milisavljevic}, {Margutti}, {Berger},
  {Calkins}, {Fong}, {Gezari}, {Huber}, {Kankare}, {Kirshner}, {Leibler},
  {Lunnan}, {Mattila}, {Marion}, {Narayan}, {Riess}, {Roth}, {Scolnic},
  {Smartt}, {Tonry}, {Burgett}, {Chambers}, {Hodapp}, {Jedicke}, {Kaiser},
  {Magnier}, {Metcalfe}, {Morgan}, {Price}, \& {Waters}}]{DCS14a}
{Drout}, M.~R. {et al.}\  2014, \apj, 794, 23

\bibitem[{{Edgar}(2004)}]{Edg04a}
{Edgar}, R. 2004, New Astron. Rev., 48, 843

\bibitem[{{Fern{\'a}ndez} \& {Metzger}(2013)}]{FM13a}
{Fern{\'a}ndez}, R. \& {Metzger}, B.~D. 2013, \mnras, 435, 502

\bibitem[{{Fern{\'a}ndez} {et~al.}(2017){Fern{\'a}ndez}, {Quataert},
  {Kashiyama}, \& {Coughlin}}]{FQK17a}
{Fern{\'a}ndez}, R., {Quataert}, E., {Kashiyama}, K., \& {Coughlin}, E.~R.
  2017, ArXiv e-prints: 1710.01735

\bibitem[{{Fujii} {et~al.}(2017){Fujii}, {Tanikawa}, \& {Makino}}]{FTM17a}
{Fujii}, M., {Tanikawa}, A., \& {Makino}, J. 2017, ArXiv e-prints: 1709.02058

\bibitem[{{Hagino} {et~al.}(2015){Hagino}, {Odaka}, {Done}, {Gandhi},
  {Watanabe}, {Sako}, \& {Takahashi}}]{HOD15a}
{Hagino}, K., {Odaka}, H., {Done}, C., {Gandhi}, P., {Watanabe}, S., {Sako},
  M., \& {Takahashi}, T. 2015, \mnras, 446, 663

\bibitem[{{Hashizume} {et~al.}(2015){Hashizume}, {Ohsuga}, {Kawashima}, \&
  {Tanaka}}]{HOK15a}
{Hashizume}, K., {Ohsuga}, K., {Kawashima}, T., \& {Tanaka}, M. 2015, \pasj,
  67, 58

\bibitem[{{Hodapp} {et~al.}(2004){Hodapp}, {Kaiser}, {Aussel}, {Burgett},
  {Chambers}, {Chun}, {Dombeck}, {Douglas}, {Hafner}, {Heasley}, {Hoblitt},
  {Hude}, {Isani}, {Jedicke}, {Jewitt}, {Laux}, {Luppino}, {Lupton}, {Maberry},
  {Magnier}, {Mannery}, {Monet}, {Morgan}, {Onaka}, {Price}, {Ryan},
  {Siegmund}, {Szapudi}, {Tonry}, {Wainscoat}, \& {Waterson}}]{Pan-STARRS04a}
{Hodapp}, K.~W. {et al.}\  2004, Astronomische Nachrichten, 325, 636

\bibitem[{{Hoyle} \& {Lyttleton}(1939)}]{HL39a}
{Hoyle}, F. \& {Lyttleton}, R.~A. 1939, Proceedings of the Cambridge
  Philosophical Society, 35, 405

\bibitem[{{Huarte-Espinosa} {et~al.}(2013){Huarte-Espinosa},
  {Carroll-Nellenback}, {Nordhaus}, {Frank}, \& {Blackman}}]{HCN13a}
{Huarte-Espinosa}, M., {Carroll-Nellenback}, J., {Nordhaus}, J., {Frank}, A.,
  \& {Blackman}, E.~G. 2013, \mnras, 433, 295

\bibitem[{{Inayoshi} {et~al.}(2017){Inayoshi}, {Hirai}, {Kinugawa}, \&
  {Hotokezaka}}]{IHK17a}
{Inayoshi}, K., {Hirai}, R., {Kinugawa}, T., \& {Hotokezaka}, K. 2017, \mnras,
  468, 5020

\bibitem[{{Inayoshi} {et~al.}(2016){Inayoshi}, {Kashiyama}, {Visbal}, \&
  {Haiman}}]{IKV16a}
{Inayoshi}, K., {Kashiyama}, K., {Visbal}, E., \& {Haiman}, Z. 2016, \mnras,
  461, 2722

\bibitem[{{Jiang} {et~al.}(2014){Jiang}, {Stone}, \& {Davis}}]{JSD14a}
{Jiang}, Y.-F., {Stone}, J.~M., \& {Davis}, S.~W. 2014, \apj, 796, 106

\bibitem[{{Kasen} \& {Bildsten}(2010)}]{KB10a}
{Kasen}, D. \& {Bildsten}, L. 2010, \apj, 717, 245

\bibitem[{{Kasen} {et~al.}(2015){Kasen}, {Fern{\'a}ndez}, \& {Metzger}}]{KFM15a}
{Kasen}, D., {Fern{\'a}ndez}, R.,  \& {Metzger}, B.~D. 2015, \mnras, 450, 1777

\bibitem[{{Kashiyama} \& {Quataert}(2015)}]{KQ15a}
{Kashiyama}, K. \& {Quataert}, E. 2015, \mnras, 451, 2656

\bibitem[{{Kimura} {et~al.}(2017{\natexlab{a}}){Kimura}, {Murase}, \&
  {M{\'e}sz{\'a}ros}}]{KMM17a}
{Kimura}, S.~S., {Murase}, K., \& {M{\'e}sz{\'a}ros}, P. 2017{\natexlab{a}},
  ArXiv e-prints: 1702.07337

\bibitem[{{Kimura} {et~al.}(2017{\natexlab{b}}){Kimura}, {Takahashi}, \&
  {Toma}}]{KTT17a}
{Kimura}, S.~S., {Takahashi}, S.~Z., \& {Toma}, K. 2017{\natexlab{b}}, \mnras,
  465, 4406

\bibitem[{{Kimura} {et~al.}(2014){Kimura}, {Toma}, \& {Takahara}}]{ktt14}
{Kimura}, S.~S., {Toma}, K., \& {Takahara}, F. 2014, \apj, 791, 100

\bibitem[{{Kinugawa} {et~al.}(2014){Kinugawa}, {Inayoshi}, {Hotokezaka},
  {Nakauchi}, \& {Nakamura}}]{KIH14a}
{Kinugawa}, T., {Inayoshi}, K., {Hotokezaka}, K., {Nakauchi}, D., \&
  {Nakamura}, T. 2014, \mnras, 442, 2963

\bibitem[{{Kiuchi} {et~al.}(2015){Kuchi}, {Sekiguchi}, {Kyutoku}, {Shibata}, {Taniguchi}, \& {Wada}}]{KSK15a}
{Kiuchi}, K., {Sekiguchi}, Y., \& {Kyutoku}, K., Shibata, M., Taniguchi, K., Wada, T. 2015, Physical Review D, 92, 064034

\bibitem[{{Komissarov}(2004)}]{Kom04a}
{Komissarov}, S.~S. 2004, \mnras, 350, 427

\bibitem[{{Kushnir} {et~al.}(2016){Kushnir}, {Zaldarriaga}, {Kollmeier}, \&
  {Waldman}}]{KZK16a}
{Kushnir}, D., {Zaldarriaga}, M., {Kollmeier}, J.~A., \& {Waldman}, R. 2016,
  \mnras, 462, 844

\bibitem[{{Law} {et~al.}(2009){Law}, {Kulkarni}, {Dekany}, {Ofek}, {Quimby},
  {Nugent}, {Surace}, {Grillmair}, {Bloom}, {Kasliwal}, {Bildsten}, {Brown},
  {Cenko}, {Ciardi}, {Croner}, {Djorgovski}, {van Eyken}, {Filippenko}, {Fox},
  {Gal-Yam}, {Hale}, {Hamam}, {Helou}, {Henning}, {Howell}, {Jacobsen},
  {Laher}, {Mattingly}, {McKenna}, {Pickles}, {Poznanski}, {Rahmer}, {Rau},
  {Rosing}, {Shara}, {Smith}, {Starr}, {Sullivan}, {Velur}, {Walters}, \&
  {Zolkower}}]{PTF09a}
{Law}, N.~M. {et al.}\  2009, \pasp, 121, 1395

\bibitem[{{Li} \& {Paczy{\'n}ski}(1998)}]{LP98a}
{Li}, L.-X. \& {Paczy{\'n}ski}, B. 1998, \apjl, 507, L59

\bibitem[{{Lovegrove} \& {Woosley}(2013)}]{LW13a}
{Lovegrove}, E. \& {Woosley}, S.~E. 2013, \apj, 769, 109

\bibitem[{{LSST Science Collaboration} {et~al.}(2009){LSST Science
  Collaboration}, {Abell}, {Allison}, {Anderson}, {Andrew}, {Angel}, {Armus},
  {Arnett}, {Asztalos}, {Axelrod}, \& et~al.}]{LSST09a}
{LSST Science Collaboration} {et al.}\  2009, ArXiv e-prints: 0912.0201

\bibitem[{{MacFadyen} \& {Woosley}(1999)}]{MW99a}
{MacFadyen}, A.~I. \& {Woosley}, S.~E. 1999, \apj, 524, 262

\bibitem[{{Mandel} \& {de Mink}(2016)}]{MM16a}
{Mandel}, I. \& {de Mink}, S.~E. 2016, \mnras, 458, 2634

\bibitem[{{Marchant} {et~al.}(2016){Marchant}, {Langer}, {Podsiadlowski},
  {Tauris}, \& {Moriya}}]{MLP16a}
{Marchant}, P., {Langer}, N., {Podsiadlowski}, P., {Tauris}, T.~M., \&
  {Moriya}, T.~J. 2016, \aap, 588, A50

\bibitem[{{Matzner} \& {McKee}(1999)}]{MM99a}
{Matzner}, C.~D. \& {McKee}, C.~F. 1999, \apj, 510, 379

\bibitem[{{Merloni} {et~al.}(2012){Merloni}, {Predehl}, {Becker},
  {B{\"o}hringer}, {Boller}, {Brunner}, {Brusa}, {Dennerl}, {Freyberg},
  {Friedrich}, {Georgakakis}, {Haberl}, {Hasinger}, {Meidinger}, {Mohr},
  {Nandra}, {Rau}, {Reiprich}, {Robrade}, {Salvato}, {Santangelo}, {Sasaki},
  {Schwope}, {Wilms}, \& {German eROSITA Consortium}}]{eROSITA12a}
{Merloni}, A. {et al.}\  2012, ArXiv e-prints: 1209.3114

\bibitem[{{Metzger} {et~al.}(2014){Metzger}, {Vurm}, {Hasco{\"e}t}, \&
  {Beloborodov}}]{MVH14a}
{Metzger}, B.~D., {Vurm}, I., {Hasco{\"e}t}, R., \& {Beloborodov}, A.~M. 2014,
  \mnras, 437, 703

\bibitem[{{Moriya} {et~al.}(2010){Moriya}, {Tominaga}, {Tanaka}, {Nomoto},
  {Sauer}, {Mazzali}, {Maeda}, \& {Suzuki}}]{MTT10a}
{Moriya}, T., {Tominaga}, N., {Tanaka}, M., {Nomoto}, K., {Sauer}, D.~N.,
  {Mazzali}, P.~A., {Maeda}, K., \& {Suzuki}, T. 2010, \apj, 719, 1445

\bibitem[{{Morokuma} {et~al.}(2014){Morokuma}, {Tominaga}, {Tanaka}, {Mori},
  {Matsumoto}, {Kikuchi}, {Shibata}, {Sako}, {Aoki}, {Doi}, {Kobayashi},
  {Maehara}, {Matsunaga}, {Mito}, {Miyata}, {Nakada}, {Soyano}, {Tarusawa},
  {Miyazaki}, {Nakata}, {Okada}, {Sarugaku}, {Richmond}, {Akitaya}, {Aldering},
  {Arimatsu}, {Contreras}, {Horiuchi}, {Hsiao}, {Itoh}, {Iwata}, {Kawabata},
  {Kawai}, {Kitagawa}, {Kokubo}, {Kuroda}, {Mazzali}, {Misawa}, {Moritani},
  {Morrell}, {Okamoto}, {Pavlyuk}, {Phillips}, {Pian}, {Sahu}, {Saito}, {Sano},
  {Stritzinger}, {Tachibana}, {Taddia}, {Takaki}, {Tateuchi}, {Tomita},
  {Tsvetkov}, {Ui}, {Ukita}, {Urata}, {Walker}, \& {Yoshii}}]{KISS14a}
{Morokuma}, T. {et al.}\  2014, \pasj, 66, 114

\bibitem[{{Murase} {et~al.}(2015){Murase}, {Kashiyama}, {Kiuchi}, \&
  {Bartos}}]{MKK15a}
{Murase}, K., {Kashiyama}, K., {Kiuchi}, K., \& {Bartos}, I. 2015, \apj, 805,
  82

\bibitem[{{Murase} {et~al.}(2016){Murase}, {Kashiyama}, {M{\'e}sz{\'a}ros},
  {Shoemaker}, \& {Senno}}]{MKM16a}
{Murase}, K., {Kashiyama}, K., {M{\'e}sz{\'a}ros}, P., {Shoemaker}, I., \&
  {Senno}, N. 2016, \apjl, 822, L9

\bibitem[{{Nadezhin}(1980)}]{Nad80a}
{Nadezhin}, D.~K. 1980, \apss, 69, 115

\bibitem[{{Nakamura} {et~al.}(1997){Nakamura}, {Sasaki}, {Tanaka}, \&
  {Thorne}}]{NST97a}
{Nakamura}, T., {Sasaki}, M., {Tanaka}, T., \& {Thorne}, K.~S. 1997, \apjl,
  487, L139

\bibitem[{{Narayan} {et~al.}(2017){Narayan}, {S{\c a}dowski}, \&
  {Soria}}]{NSS17a}
{Narayan}, R., {S{\c a}dowski}, A., \& {Soria}, R. 2017, \mnras, 469, 2997

\bibitem[{{Narayan} \& {Yi}(1994)}]{ny94}
{Narayan}, R. \& {Yi}, I. 1994, \apjl, 428, L13

\bibitem[{{Ohsuga} {et~al.}(2005){Ohsuga}, {Mori}, {Nakamoto}, \&
  {Mineshige}}]{OMN05a}
{Ohsuga}, K., {Mori}, M., {Nakamoto}, T., \& {Mineshige}, S. 2005, \apj, 628,
  368

\bibitem[{{Price} \& {Rosswog}(2006)}]{PR06a}
{Price}, D.~J., \& {Rosswog}, S. 2006, Science, 312, 719

\bibitem[{{Perna} {et~al.}(2014){Perna}, {Duffell}, {Cantiello}, \&
  {MacFadyen}}]{PDC14a}
{Perna}, R., {Duffell}, P., {Cantiello}, M., \& {MacFadyen}, A.~I. 2014, \apj,
  781, 119

\bibitem[{{Perna} {et~al.}(2016){Perna}, {Lazzati}, \& {Giacomazzo}}]{PLG16a}
{Perna}, R., {Lazzati}, D., \& {Giacomazzo}, B. 2016, \apjl, 821, L18

\bibitem[{{S{\c a}dowski} {et~al.}(2014){S{\c a}dowski}, {Narayan}, {McKinney},
  \& {Tchekhovskoy}}]{SNM14a}
{S{\c a}dowski}, A., {Narayan}, R., {McKinney}, J.~C., \& {Tchekhovskoy}, A.
  2014, \mnras, 439, 503

\bibitem[{{Schaerer} \& {Maeder}(1992)}]{SM92a}
{Schaerer}, D. \& {Maeder}, A. 1992, \aap, 263, 129

\bibitem[{{Shima} {et~al.}(1985){Shima}, {Matsuda}, {Takeda}, \&
  {Sawada}}]{SMT85a}
{Shima}, E., {Matsuda}, T., {Takeda}, H., \& {Sawada}, K. 1985, \mnras, 217,
  367

\bibitem[{{Sigurdsson} \& {Hernquist}(1993)}]{SH93a}
{Sigurdsson}, S. \& {Hernquist}, L. 1993, \nat, 364, 423

\bibitem[{{Smith} {et~al.}(2011){Smith}, {Li}, {Silverman}, {Ganeshalingam}, \&
  {Filippenko}}]{SLS11a}
{Smith}, N., {Li}, W., {Silverman}, J.~M., {Ganeshalingam}, M., \&
  {Filippenko}, A.~V. 2011, \mnras, 415, 773

\bibitem[{{Takahara} \& {Kusunose}(1985)}]{tk85}
{Takahara}, F. \& {Kusunose}, M. 1985, Progress of Theoretical Physics, 73,
  1390

\bibitem[{{Takahashi} \& {Ohsuga}(2015)}]{TO15a}
{Takahashi}, H.~R. \& {Ohsuga}, K. 2015, \pasj, 67, 60

\bibitem[{{Takahashi} {et~al.}(2016){Takahashi}, {Ohsuga}, {Kawashima}, \&
  {Sekiguchi}}]{TOK16a}
{Takahashi}, H.~R., {Ohsuga}, K., {Kawashima}, T., \& {Sekiguchi}, Y. 2016,
  \apj, 826, 23

\bibitem[{{Tanaka} {et~al.}(2016){Tanaka}, {Tominaga}, {Morokuma}, {Yasuda},
  {Furusawa}, {Baklanov}, {Blinnikov}, {Moriya}, {Doi}, {Jiang}, {Kato},
  {Kikuchi}, {Kuncarayakti}, {Nagao}, {Nomoto}, \& {Taniguchi}}]{TTM16b}
{Tanaka}, M. {et al.}\  2016, \apj, 819, 5

\bibitem[{{Tassoul}(1987)}]{Tas87a}
{Tassoul}, J.-L. 1987, \apj, 322, 856

\bibitem[{{Toma} \& {Takahara}(2016)}]{TT16a}
{Toma}, K. \& {Takahara}, F. 2016, Progress of Theoretical and Experimental
  Physics, 2016, 063E01

\bibitem[{{Tutukov} \& {Yungelson}(1993)}]{TY93a}
{Tutukov}, A.~V. \& {Yungelson}, L.~R. 1993, \mnras, 260, 675

\bibitem[{{Yoon} {et~al.}(2010){Yoon}, {Woosley}, \& {Langer}}]{YWL10a}
{Yoon}, S.-C., {Woosley}, S.~E., \& {Langer}, N. 2010, \apj, 725, 940

\bibitem[{{Zahn}(1977)}]{Zah77a}
{Zahn}, J.-P. 1977, \aap, 57, 383

\end{thebibliography}

\listofchanges

\end{document}